\documentclass[12pt,noshowpacs,nofootinbib,notitlepage,amsmath]{revtex4-1}
\pdfoutput=1
\usepackage{setspace}
\allowdisplaybreaks
\usepackage{graphicx,color,amssymb}
\usepackage[colorlinks=true,citecolor=blue,linkcolor=blue,urlcolor=blue]{hyperref}
\usepackage[charter]{mathdesign}
\DeclareSymbolFontAlphabet{\mathcal}{symbols}
\DeclareSymbolFont{symbols}{OMS}{xmdcmsy}{m}{n}
\DeclareSymbolFont{largesymbols}{OMX}{xmdcmex}{m}{n}
\SetSymbolFont{symbols}{bold}{OMS}{xmdcmsy}{b}{n}

\def\Mp{M_{\mathrm{Pl}}}

\begin{document}  
\title{\color{blue}\Large A QCD analogy for quantum gravity}
\author{Bob Holdom}
\email{bob.holdom@utoronto.ca}
\author{Jing Ren}
\email{jren@physics.utoronto.ca}
\affiliation{Department of Physics, University of Toronto, Toronto, Ontario, Canada  M5S1A7}
\begin{abstract}
Quadratic gravity presents us with a renormalizable, asymptotically free theory of quantum gravity. When its couplings grow strong at some scale, as in QCD, then this strong scale sets the Planck mass. QCD has a gluon that does not appear in the physical spectrum. Quadratic gravity has a spin-2 ghost that we conjecture does not appear in the physical spectrum. We discuss how the QCD analogy leads to this conjecture and to the possible emergence of general relativity. Certain aspects of the QCD path integral and its measure are also similar for quadratic gravity. With the addition of the Einstein-Hilbert term, quadratic gravity has a dimensionful parameter that seems to control a quantum phase transition and the size of a mass gap in the strong phase.
\end{abstract}
\maketitle

\section{Introduction}
\label{intro}

With increasing precision in various measurements and observations, the general theory of relativity still successfully describes all known gravitational phenomena. While there is as yet no direct evidence for a quantum description of the gravitational field, there has been a decades-long effort to unify general relativity with quantum mechanics. The most naive attempt is to quantize general relativity in a way similar to other gauge field theories. Unfortunately it turns out to be nonrenormalizable. Additional terms with higher mass dimension are generated by quantum loops. Very early it was realized that in principle these terms can be incorporated into the action \cite{Weyl}. From a modern point of view there is no problem to treat general relativity as an effective quantum description below the Planck mass $\Mp$ \cite{Donoghue:1994dn}\cite{Burgess:2003jk}.

There are three operators that are quadratic in the curvature: $R^2$, $R^{\mu\nu}R_{\mu\nu}$, $R^{\mu\nu\alpha\beta}R_{\mu\nu\alpha\beta}$. In four dimensions the last term can be eliminated due to the Gauss-Bonnet topological invariant. Including the Einstein-Hilbert term, it is convenient to organize the quadratic action as follows
\begin{eqnarray}\label{eq:quadratic}
S_{\mathrm{QG}}=\int d^4x\,\sqrt{-g}\left(\frac{1}{2}M^2R-\frac{1}{2 f_2^2}C_{\mu\nu\alpha\beta}C^{\mu\nu\alpha\beta}+\frac{1}{3 f_0^2}R^2\right),\end{eqnarray}
with $\frac{1}{2}C_{\mu\nu\alpha\beta}C^{\mu\nu\alpha\beta}=R_{\mu\nu}R^{\mu\nu}-\frac{1}{3}R^2$ up to the topological term.\footnote{For gravity conventions we use the signature $(+\,-\,-\,-)$. The Riemann tensor is defined as $R_{\,\kappa\mu\nu}^\lambda=\partial_{\nu}\Gamma_{\kappa\mu}^{\lambda}+\Gamma_{\nu\sigma}^{\lambda}\Gamma_{\kappa\mu}^{\sigma}-(\mu\leftrightarrow\nu)$, and the Ricci tensor is defined as $R_{\kappa\nu}=R_{\,\kappa\lambda\nu}^\lambda$.} The $R$ and $R^2$ terms break the conformal gauge symmetry of the second term, the Weyl tensor term. The action is characterized by two dimensionless couplings and one mass scale. The mass scale $M$ breaks the classical scale invariance softly. In our later discussion $M$ shall not be identified with the Planck mass.

When (\ref{eq:quadratic}) is considered as a fundamental theory, rather than the first few terms of a derivative expansion, it was found that this theory is perturbatively renormalizable \cite{Stelle:1976gc}\cite{Voronov:1984kq}. The reason comes from the dominance of the higher-derivative terms in the UV. Expanding around the flat background $g_{\mu\nu}=\eta_{\mu\nu}+h_{\mu\nu}$, the gauge-fixed propagator for the metric fluctuation $h_{\mu\nu}$ tends to $1/k^4$ in the UV times tensor structures. Renormalizability is also related to how the classical scale invariance is only broken softly and by the trace anomaly. The latter corresponds to the logarithmic running of the two dimensionless couplings.\footnote{The Gauss-Bonnet topological term also plays a role in the renormalizability of the theory, but it does not contribute to the renormalization group running of these two couplings~\cite{GBterm1}\cite{GBterm2}.}

It is convenient to consider the running of $f_2^2$ and the ratio of couplings $w=f_2^2/f_0^2$. The one-loop beta functions are the following \cite{Fradkin:1981iu}\cite{Avramidi:1985ki}:
\begin{eqnarray}
\frac{d f_2^2}{dt}=-\left(\frac{133}{10}+a_m\right) f_2^4,\quad
\frac{1}{f_2^2}\frac{d w^2}{dt}=-\left[\frac{5}{12}+w\left(5+\frac{133}{10}+a_m\right)+\frac{10}{3}w^2\right]\,,
\end{eqnarray}
where $a_m>0$ denotes the matter contribution. With the same sign contributions from gravity and matter, $f_2^2>0$ is always asymptotically free. For the coupling ratio $w$, there are two roots $w_2<w_1<0$ of the beta function. The coupling ratio approaches the UV fixed point $w_1$ for $w$ within the UV attractive region $w_2<w<0$ ($w_2=-5.5, w_1=-0.023$ when ignoring $a_m$). $w\to0$ denotes a strong coupling limit where the one-loop analysis is not reliable. So the quadratic action in (\ref{eq:quadratic}) is asymptotically free for $f_2^2>0, f_0^2<0$. The magnitudes of the couplings will grow into the IR at least down to the masses of the massive gravitational modes, which are determined by $M$ and the sizes of the couplings. If the couplings are weak on the scale of the masses then the theory remains perturbative. The theory is usually considered in this perturbative phase where $M$ can be identified with the Planck mass $\Mp$.

However the nice UV behavior enjoyed by this higher-derivative gravity theory comes at a great cost, since a ghost appears in the spectrum.\footnote{This is also true of the recent proposals to set $M=0$ and to instead introduce a scalar field with nonminimal gravitational coupling. $\Mp^2R$ can then be generated perturbatively by the Coleman-Weinberg mechanism~\cite{Agravity1}\cite{Agravity2}.} The problem shows up at the classical level in the Arnowitt-Deser-Misner formalism~\cite{ADM} where it manifests itself as the Ostrogradski instability~\cite{Ostrogradski}\cite{Woodard:2015zca}. This occurs for a Hamiltonian that has a linear dependence on some canonical variables, since this implies unboundedness from below. This happens in a nondegenerate higher-derivative theory, where the highest derivative term can be expressed as a function of other canonical variables. For quadratic gravity this is the case when the Weyl tensor term is present. On the other hand the Weyl term is needed to produce the better-behaved $1/k^4$ UV behavior in the spin-2 sector, as needed for renormalizability.

The problem is more directly apparent from the propagator of $h_{\mu\nu}$ on a flat background. The tensor structure projects out spin 2, 1 and 0 degrees of freedom. The propagator in the spin-2 sector exhibits both a massless pole for the normal graviton as well as a massive pole at $M_2^2=\frac{1}{2}f_2^2M^2$ with negative residue. Depending on the prescription of the $i\epsilon$ term there are two interpretations of this massive ghost pole, as either a state of negative-norm or a state of negative energy. The first interpretation was useful to prove renormalizability~\cite{Stelle:1976gc}, but a negative norm state is difficult to reconcile with a probability interpretation and unitarity. In the negative-energy interpretation there is a vacuum instability. A negative-energy ghost that couples to positive-energy particles implies an infinite phase space for vacuum decay into collections of ghosts and normal particles. For the spin-0 sector, the additional massive pole has positive residue, but the asymptotically free condition $f_0^2<0$ means that it describes a tachyon with mass $M_0^2=\frac{1}{4}f_0^2M^2$.

At face value the ghost problem prevents quadratic gravity from being a possible UV completion of general relativity. But this conclusion is based on the analysis of a classical Hamiltonian or on the structure of tree-level propagators. Basically for any physical process that explicitly manifests the ghost problem, there is usually the implicit assumption that the perturbative analysis reflects the true physical spectrum. This will be correct when $M$ is sufficiently large, since as we have mentioned, the theory remains perturbative. But if $M$ is sufficiently small then the running couplings will grow strong at some scale $\Lambda_{\mathrm{QG}}>M$.  Then the poles appearing in the perturbative propagators fall into the nonperturbative region and the arguments based on perturbative modes deserve further consideration.

It is very instructive to consider another renormalizable and asymptotically free theory, quantum chromodynamics (QCD). It has no analog of a parameter like $M$ in the action that can prevent the coupling from growing strong, and in this case we are used to the fact that the physical spectrum bears no relation to the perturbative degrees of freedom. In particular the gluon is not in the physical spectrum. This is a consequence of confinement, but it is also understood more directly in terms of the behavior of the full gluon propagator. Essentially there is an IR suppression of the propagator that is sufficient to remove the gluon pole.

It should be noted that when calculations are done in perturbative QCD, the processes being described involve high virtuality and the gluon propagator need only describe far off-shell gluons. In this way there is a factorization; the calculation of the hard process in a high-energy scattering event is independent of the IR physics that is responsible for removing the gluon from the spectrum. With this understanding the calculation of the hard process can be performed using the tree level gluon propagator. The gluon loosely speaking propagates as a virtual particle but not as a physical on-shell particle. The theory has only one physical spectrum and it does not include a gluon. Similarly the full gluon propagator describes both large- and small-$|k^2|$ behavior and it does not have a pole. The four-momentum can have arbitrarily large components, but if $|k^2|$ is small then it probes the nonperturbative IR behavior of this propagator.

We propose here that a similar story holds for the full graviton propagator, and that it is modified in the IR in such a way that the spin-2 ghost pole is absent. Then the ghost is not in the physical spectrum, since the existence and location of poles in the full propagator are statements about the physical spectrum even though the full propagator displays gauge dependence. With regard to the vacuum instability mentioned above, this instability can only occur if negative-energy states do in fact exist in the physical spectrum. Thus we propose that the theory enters a distinctly different phase when the parameter $M$ falls below some critical value $\sim\Lambda_{\mathrm{QG}}$, with a distinctly different physical spectrum in this strong phase. Both the spin-2 ghost and the spin-0 tachyon could be absent, or in the case of the tachyon, large quantum corrections could instead just change the sign of its mass.

We shall argue that $M$ controls the size of a mass gap for the graviton, and that it is only for $M=0$ that a normal massless graviton could emerge in the IR. In that case general relativity becomes the effective description in the IR, and $\Mp$ is identified with the strong gravity scale $\Lambda_{\mathrm{QG}}$. This emergence of the spin-2 massless graviton is not in contradiction with the Weinberg-Witten theorem \cite{WW} since the fundamental theory is diffeomorphism invariant in the same sense as general relativity. For nonzero $M$ we shall discuss different possibilities for how the graviton mass gap is realized, based on analogies with either dynamical symmetry breaking or confinement. In addition we argue that $M$ gives control over a quantum phase transition between the weak and strong phases as $M$ moves below $\Lambda_{\mathrm{QG}}$. This is a feature not present in QCD, and it is interesting to consider the nature of this phase transition.

One may wonder about how a consistent quantum theory can emerge when the action is problematic at tree level. We shall use a path integral as a nonperturbative definition of the quantum theory, and with this there are explicit effects that can alter the tree-level analysis. A gauge theory path integral has a nontrivial measure that is constructed to uniformly sample the gauge orbits in configuration space. A similar construction applies to both QCD and gravity. In QCD it is known that the measure brings in effects related to the nontrivial structure of the gauge configuration space, effects that make the proper sampling of gauge orbits difficult to attain (Gribov copies \cite{Gribov:1977wm}). These nonlocal, nonperturbative, but nondynamical effects (effects explicitly present in the definition of the theory) are thought to be themselves capable of the suppression and the removal of the gluon.

Some speculation about an analogy between quadratic gravity and QCD (which extends an old analogy between general relativity and the chiral Lagrangian of QCD) has occurred before \cite{analogy}\cite{Maggiore:2015rma}. However with the help of a more recent understanding of nonperturbative QCD, we hope to provide a more consistent picture of the analogy, including the fate of the ghost, the emergence of general relativity, the identification of $\Mp$, the effect of $M$ and so on. We also highlight similarities in the nonperturbative path integral definitions of the two theories. Presently we lack more direct arguments as to why the analogy should hold. Rather the analogy provides some initial expectations for the behavior of strong quadratic gravity, and it counters the negative sentiment towards quadratic gravity that is based solely on the weak (large-$M$) theory.

We briefly comment on various other proposals to solve the ghost problem. Tomboulis considered the loop effects of $N$ fermions in the large-$N$ limit and found that the ghost pole was transformed into a pair of complex-conjugate poles~\cite{Tomboulis:1977jk}, which requires an implementation of the Lee-Wick prescription \cite{leewick} for quadratic gravity. Whether unitarity is achieved remains unsettled~\cite{Antoniadis:1986tu}\cite{Johnston}. There are various studies of toy models of higher-derivative theories but it remains doubtful that the ghost problem can be solved at the perturbative level~\cite{Pais:1950za}\cite{ToyModel1}\cite{ToyModel2}\cite{Woodard:2015zca}. An attempt to impose additional constraints to avoid the ghost ended up with a violation of Lorentz invariance~\cite{Chen:2013aha}. Alternative quantization schemes have been considered~\cite{Mannheim}, also very recently \cite{Salvio:2015gsi}. Other authors go beyond quadratic gravity by considering a Lorentz-violating modification in the UV~\cite{Horava:2009uw} or by proposing to sum up all higher-derivative terms to produce a ghost-free nonlocal classical action~\cite{Tomboulis1997}\cite{Modesto:2011kw}\cite{Biswas:2011ar}. A nonperturbative proposal relies on a nontrivial renormalization group running of the Planck mass in the UV \cite{RGE1}\cite{RGE3}. Similarly if the asymptotic safety scenario can be realized for gravity then it is hoped that the perturbative ghost pole is either removed or set to infinite mass \cite{RGE2}. In~\cite{Tomboulis:1983sw} it was argued that quadratic gravity could be formulated on a lattice in such a way as to be ghost-free and unitary.

The rest of the paper is organized as follows. In Sec.~\ref{sec:analogy} we develop a description of the nonperturbative effects in QCD in a way that we can simply carry over to quadratic gravity. This apparently leads to a healthy theory when $M\lesssim \Lambda_{\textrm{QG}}$. We treat the $M=0$ case first and then we go on to consider the effect of $M$ and a quantum phase transition controlled by the value of $M$. In Sec.~\ref{sec:refine} we discuss the measure of the path integral of a gauge theory as an origin of the nonperturbative effects of interest, and then find some evidence that the gravity measure is also influenced by Gribov copies. Finally, we discuss some implications and questions of this picture of gravity in Sec.~\ref{sec:implication}.

\section{Propagators and a phase transition}
\label{sec:analogy}


We start with the gluon propagator, by which we mean the full nonperturbative gluon two-point function. The propagator is of course gauge dependent, but the different gauges will have to agree on the existence or not of a massless gluon pole. For nonperturbative studies it is common to respect the spacetime symmetries by choosing a covariant gauge such as Landau gauge ($\partial^\mu A_\mu=0$). Then various approaches to nonperturbative QCD indicate that the gluon propagator is suppressed in the IR such that the massless pole is removed. A mass gap develops without a standard massive pole and without breaking the gauge symmetries, and it is this phenomenon that is now better understood. One consequence of the mass gap is that the light gluon contribution to the running gauge coupling is removed. This then also resolves the issue of an IR Landau pole that is naively implied by perturbative analysis.

The lattice studies and typically also the continuum Schwinger-Dyson studies are performed in Euclidean space, and thus only probe the gluon propagator for spacelike momenta, $k^2<0$. This is sufficient to see the absence of a pole at $k^2=0$ and it is also sufficient to detect a nonperturbative violation of positivity, by which we mean that the propagator has no representation in terms of a positive spectral density~\cite{Alkofer:2000wg}. We note though that a positive spectral density is not necessarily an expected property of a gauge field propagator. In particular through a renormalization group analysis it is found the gluon propagator falls faster than $1/k^2$, and this implies a violation of positivity in the perturbative regime \cite{Oehme:1979ai}\cite{Cornwall:2013zra}. We mention this here since in quadratic gravity the UV behavior of the propagator is $1/k^4$, which thus also violates positivity. At the tree level the violation is due to the ghost contribution, but we shall argue that a nonperturbative violation is also possible without the occurrence of a ghost.

Returning to QCD, a nonperturbative violation of positivity was observed through the study of the Schwinger-Dyson equations \cite{SDS}\cite{SDD}, although the detailed behavior in the deep IR was left open. The propagator vanished as a noninteger power as $k^2\rightarrow0$ for the ``scaling solution'', while it approached a nonvanishing value for the ``decoupling solution''. In addition to these dynamical effects there are the nondynamical aspects of nonperturbative QCD arising from Gribov copies, which is our focus in Sec.~\ref{sec:refine}. The Gribov-Zwanziger approach~\cite{Zwanziger:1989mf}\cite{RevGZ} (not our focus in Sec.~\ref{sec:refine}) predicted a vanishing propagator as $k^2\to0$, a prediction that was later refined to agree with the decoupling solution~\cite{RefineGZ}. The most reliable evidence is provided by lattice studies that implement gauge fixing, and these studies can potentially capture both the dynamical and nondynamical effects. The results in Landau gauge confirm the mass gap and in particular have given strong support for the decoupling solution~\cite{LGlattice1}\cite{LGlattice2}. 

\begin{figure}[!h]
  \centering%
{ \includegraphics[width=7.4cm]{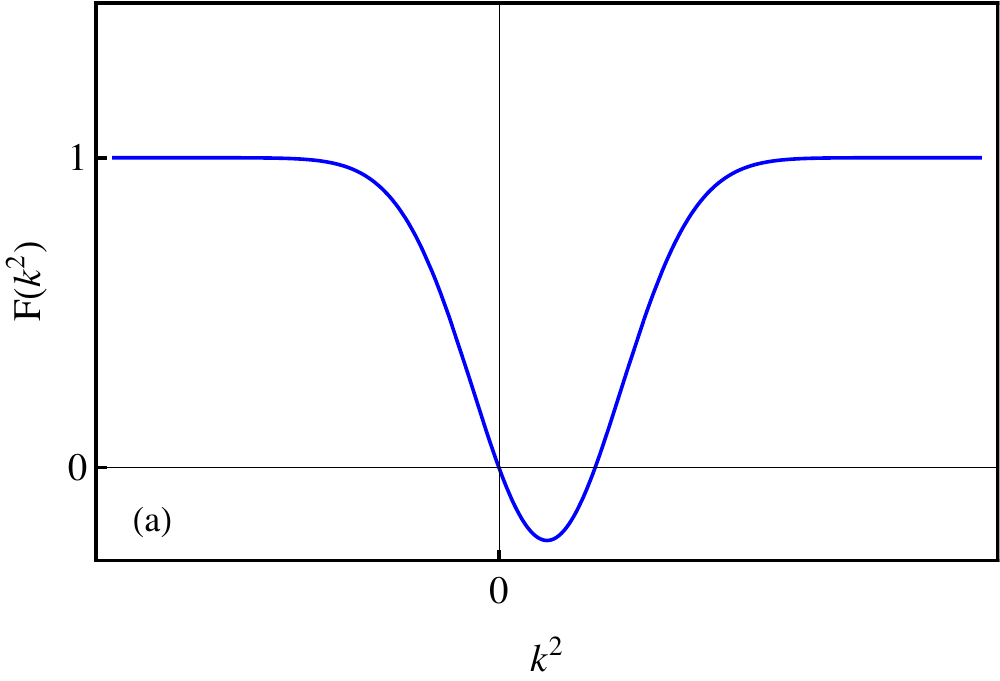}}\quad\quad
{ \includegraphics[width=7.4cm]{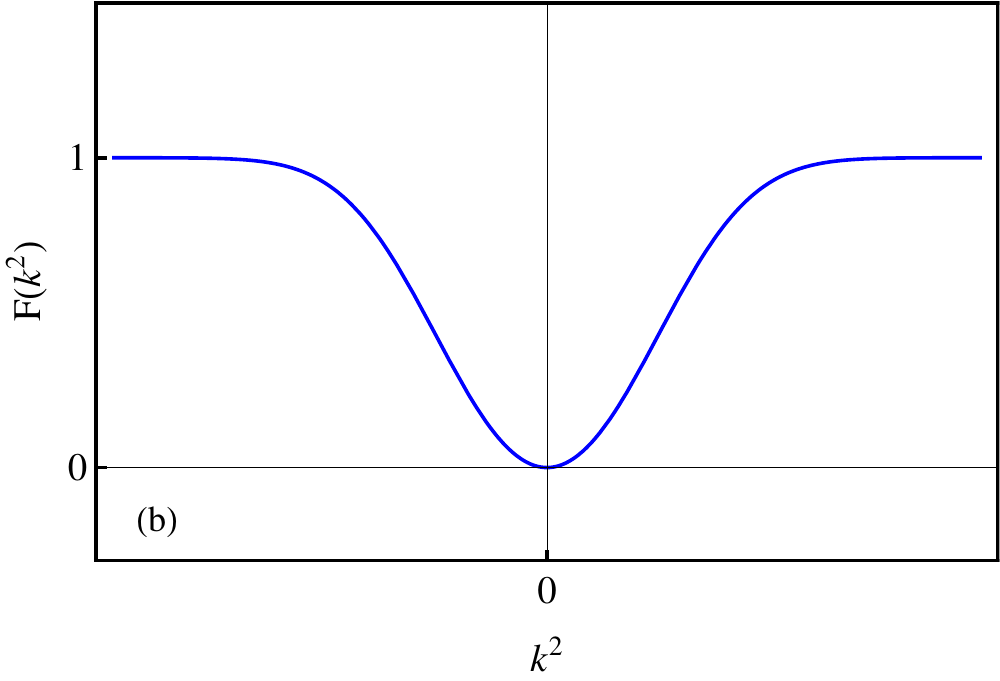}}
\caption{\label{fig:Propagator} The multiplicative factor $F(k^2)$ in the gluon propagator, in either Landau gauge (a) or Coulomb gauge (b). }
\end{figure}

We shall encode the nonperturbative effects that appear in the gluon propagator as a real multiplicative factor $F(k^2)$. Then the propagator is $F(k^2)/k^2$ times a tensor factor and a perturbative correction factor. Imaginary parts can be absorbed into this last factor. The nonperturbative effect only operates in the IR and so $F(k^2)\rightarrow1$ for $k^2\rightarrow\pm\infty$. Due to the Euclidean space evidence for the decoupling solution we require that $F(k^2)/k^2$ approaches a negative constant in the $k^2\to0^-$ limit. Then $F(0)=0$ and $F'(0)<0$. $F(k^2)$ must have another zero for $k^2>0$ (and not a pole) to recover $F(k^2)\rightarrow1$ for $k^2\rightarrow\infty$. Thus with two zeros we have a shape for $F(k^2)$ as shown in Fig.~\ref{fig:Propagator}(a). The refined Gribov-Zwanziger propagator also implies this shape for $F(k^2)$.

Another sensible gauge for nonperturbative studies is Coulomb gauge ($\partial^iA_i=0$). It maintains rotational invariance while reflecting the space and time split of a Hamiltonian formulation. The Gauss law constraint arises directly from a phase space path integral. In this gauge there is some evidence that the transverse gluon propagator vanishes for $k^2\to0$, both from lattice studies \cite{Nakagawa:2009zf} and from the original Gribov picture~\cite{Gribov:1977wm}. This would mean that $F(k^2)$ takes the shape in Fig.~\ref{fig:Propagator}(b) in this gauge. Although distinctly different from Landau gauge, the physical implications are the same. There is no physical pole and no on-shell gluon, and there is a nonperturbative violation of positivity.

Now let us see what happens if we assume that nonperturbative effects in quadratic gravity operate in a way similar to QCD. We start with $M=0$ in (\ref{eq:quadratic}). This is perturbatively stable if we assume that the classical scale invariance holds for the matter sector as well. We also focus on a flat background, in which case the strong gravity scale $\Lambda_{\mathrm{QG}}$ is the only mass scale. We again encode nonperturbative effects in terms of a factor $G(k^2)$, so that the graviton propagator in a covariant gauge is $-G(k^2)/k^4$ times tensor and perturbative correction factors. From our experience with QCD we can consider a $G(k^2)$ that takes one of two simple forms, that is the form of $F(k^2)$ in Fig.~\ref{fig:Propagator}(a) or (b).

\begin{figure}[!h]
  \centering%
{ \includegraphics[width=7.4cm]{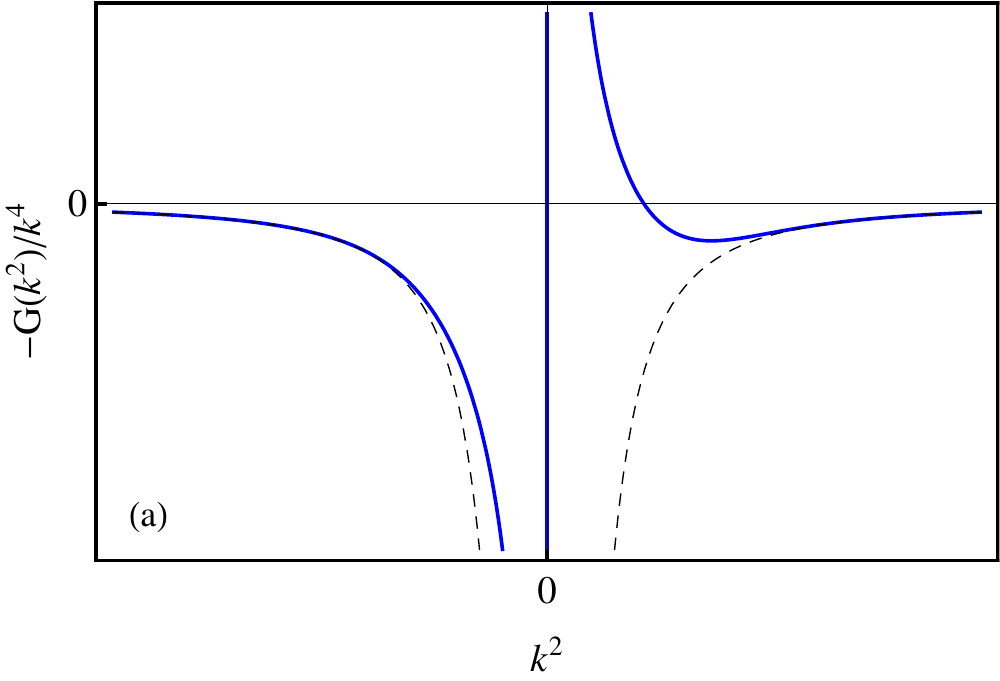}}\quad\quad
{ \includegraphics[width=7.4cm]{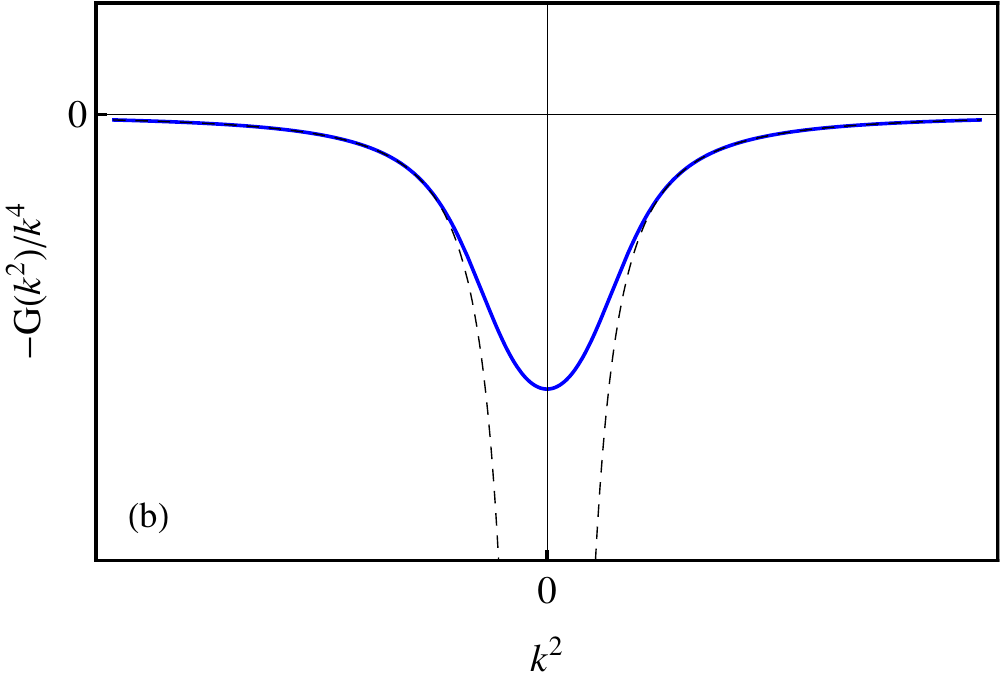}}
\caption{\label{fig:Gpropagator} The nonperturbative graviton propagator $-G(k^4)/k^4$ (blue solid line). (a) and (b) correspond to $G(k^2)$ taking the form of (a) and (b) in Fig.~\ref{fig:Propagator}. The black dashed line is $-1/k^4$.}
\end{figure}

In the first case the resulting $-G(k^2)/k^4$ is shown in Fig.~\ref{fig:Gpropagator}(a). The $-1/k^4$ behavior has been softened to $1/k^2$ with positive residue but no mass gap, and a zero now appears in the propagator at some $k^2>0$. The other choice for the form of $G(k^2)$ gives the $-G(k^2)/k^4$ as shown in Fig.~\ref{fig:Gpropagator}(b). Here a mass gap for the graviton has arisen as in QCD. In both cases the physical spectrum has changed in a way that may be consistent with a healthy theory. The two possibilities for $G(k^2)$ that we have considered are physically distinct, unlike the two cases for $F(k^2)$ that we considered for QCD. While the two forms for the gluon propagator correspond to gauge choices, the existence or not of a massless graviton is a statement about the physical spectrum.

The first case is of interest because of its implication of a massless particle with standard $1/k^2$ behavior in the spin-2 sector. If general covariance is still maintained by the strong dynamics, as we shall assume, then the complete nonlinear description of the massless spin-2 particle requires general relativity. The Planck mass is identified as
\begin{eqnarray}
\Mp^2=-1/G'(0)\sim\Lambda_{\mathrm{QG}}^2\,.
\end{eqnarray}
At energy far below $\Mp$, the standard nonrenormalizable theory with a dimensionful coupling emerges, with a weakly coupled gravitational interaction mediated by this massless graviton. The linearly rising potential $V(r)\sim r$ as implied by the $1/k^4$ propagator in the UV has made a transition to the Newtonian potential $V(r)\sim 1/r$ as implied by $1/k^2$ in the IR.

Similar to the chiral Lagrangian, the IR physics is expected to be described by a derivative expansion of the curvature tensors with a leading Einstein-Hilbert term,
\begin{eqnarray}\label{eq:derexp}
S_{\mathrm{EFT}}=\int d^4x \,\sqrt{-g}\left(\frac{1}{2}\Mp^2R+c_1R^2+c_2C_{\mu\nu\alpha\beta}C^{\mu\nu\alpha\beta}+...\right)
.\end{eqnarray}
The quadratic terms of curvature appearing here have no direct relation to those in the original quadratic gravity (\ref{eq:quadratic}). The order-one coefficients $c_1, c_2, ...$ encode the effects of strong interactions and a subset are related for example to the derivative expansion of the full inverse propagator. At any finite order the inverse propagator may have new zeros, but this does not mean new particles, since the effective theory has already broken down at the apparent new masses. Perturbative corrections in this effective theory do produce a running of the couplings of higher-derivative terms, but IR Landau poles, as naively present in the $f_0$ and $f_2$ couplings of the underlying theory, have been avoided. This picture of a theory that is weakly interacting in the UV and IR limits, with only an intermediate region that is strong, is very analogous to QCD, where in that case it is the chiral Lagrangian that describes a weak IR theory. The theory of pions also has no mass gap in the limit of vanishing current quark masses, but the difference is that for gravity it is the same field $g_{\mu\nu}$ that appears in both the UV and IR descriptions.

Now let us consider the effect of the mass parameter $M$ in the original quadratic gravity (\ref{eq:quadratic}). For very large $M$ there are the ghost and tachyon states with mass-squares $M_2^2$ and $M_0^2$ as given above. The running couplings $f_2$ and $f_0$ can be defined by their values at the scale of these masses. These couplings can be very small but as $M$ decreases these couplings increase. For larger couplings, $M_2, M_0$ can become of order $M$, but as $M$ decreases eventually a new dynamical mass scale $\Lambda_{\mathrm{QG}}$ appears. The naive perturbative masses have dropped below $\Lambda_{\mathrm{QG}}$ and we expect that a quantum phase transition has occurred to a phase with a different physical spectrum. In the following we consider two possibilities for the nature of this phase transition, with both being motivated by known phenomena in gauge theories. The first is analogous to dynamical symmetry breaking while the second is analogous to confinement.

The first possibility is suggested by a naive extension to $M\neq0$ of our previous nonperturbative modification of the spin-2 graviton propagator for $M=0$. With $G(k^2)$ behaving as in Fig.~\ref{fig:Gpropagator}(a) the only pole of $-G(k^2)/(k^2(k^2-M_2^2))$ is at $k^2=M_2^2$, and with the second zero of $G(k^2)$ located to the right of $M_2^2$, this pole has a residue of the right sign. What is originally a massive ghost pole becomes a normal massive graviton pole. The pole mass may be modified by strong interactions, but still $M_2^2\sim M^2$. A massive graviton though is not a simple concept and the required dynamics here is much less trivial than this simple argument would indicate. In fact no UV-complete theory of a massive graviton has been found~\cite{RevMassG1}\cite{RevMassG2}. A massive graviton cannot be described in a local diffeomorphism invariant way, and so the appearance of a massive graviton in a theory that is fundamentally diffeomorphism-invariant suggests a dynamical symmetry breaking of this symmetry. We see no reason not to expect a generic set of Lorentz-invariant mass terms for the metric fluctuation, rather than a set that is tuned to avoid a ghost as in the Fierz-Pauli theory. There are then six degrees of freedom, the two transverse modes and four Goldstone modes, the latter being the three longitudinal modes for the spin-2 massive graviton and the additional spin-0 ghost. Since the Goldstone modes are dynamically generated in a strongly coupled theory they need only exist up to the scale of the symmetry breaking, which is of order $M$. The appearance of the spin-0 ghost suggests that the theory is not well behaved unless one takes the $M\to0$ limit, to thereby recover our previous $M=0$ description.

One might wonder a little about the vacuum instability at $M>0$ implied by such a ghost. The question is whether the allowed values of the four-momenta of this composite particle are constrained, as for example was the case for a ghost associated with a Lorentz symmetry breaking cutoff as discussed in \cite{Cline:2003gs}. Then the vacuum decay amplitude could have a finite rather than an infinite phase space, and the vacuum instability could become controlled and possibly safe. Nevertheless there are aspects of this picture that remain quite mysterious and so we move on to the second possibility, which to us makes more sense.

\begin{figure}[!h]
  \centering%
{ \includegraphics[width=7.4cm]{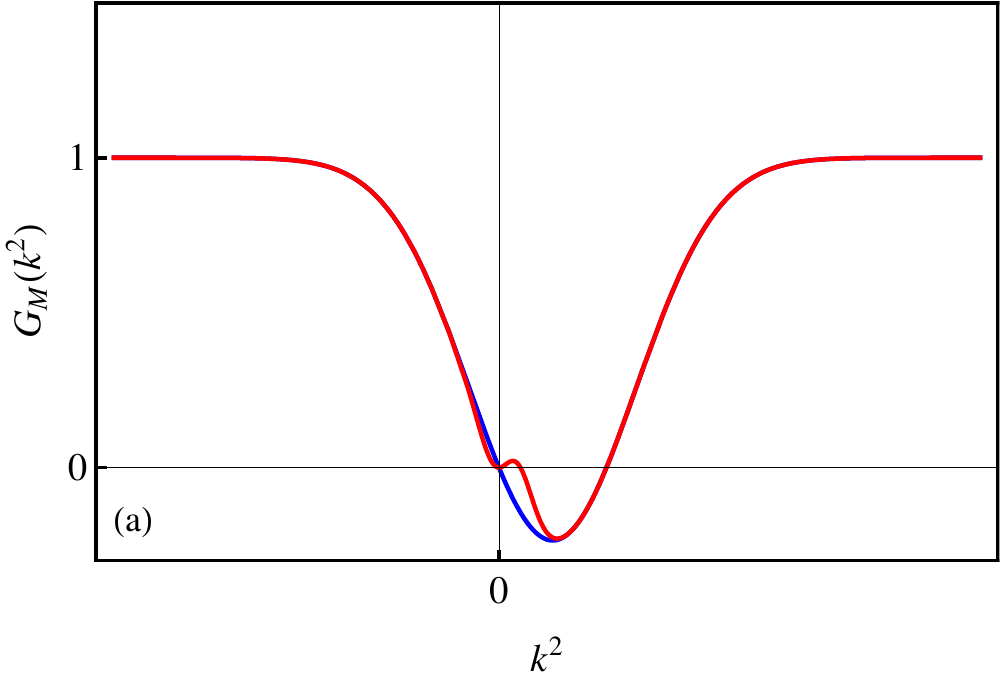}}\quad\quad
{ \includegraphics[width=7.4cm]{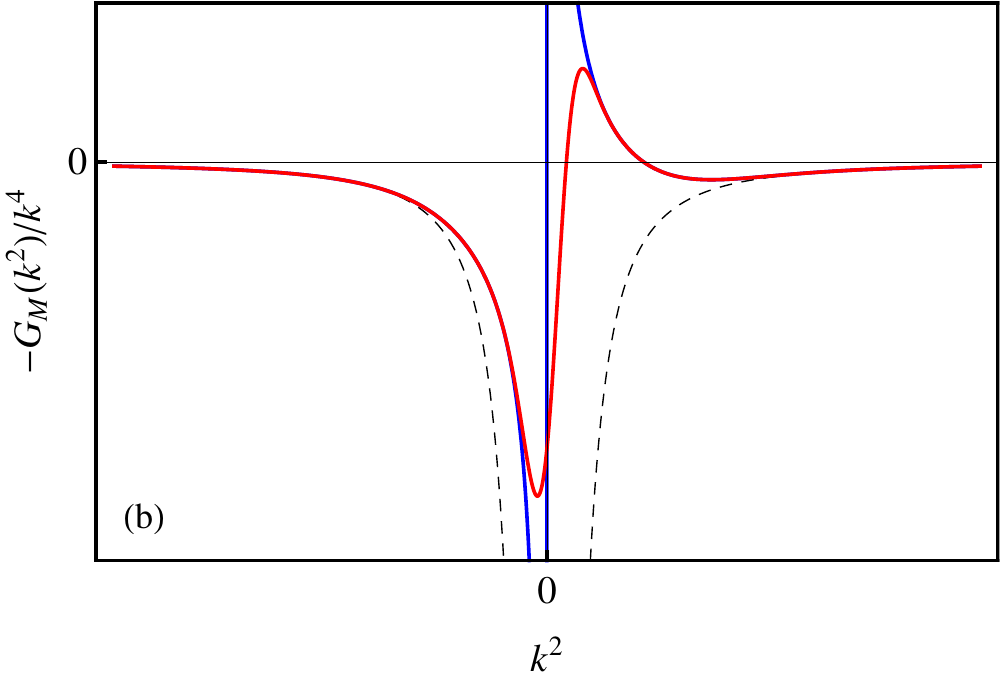}}
\caption{\label{fig:GMpropagator} The factor $G_M(k^2)$ (a) and nonperturbative graviton propagator $-G_M(k^2)/k^4$ (b) when $M\neq0$ (red solid line). The blue line has $M=0$ and the black dashed line is $-1/k^4$.}
\end{figure}

This other possibility for the quantum phase transition seems more analogous to QCD. As $M$ drops below $\Lambda_{\mathrm{QG}}\sim\Mp$ we can consider that the theory enters a confining phase. By this we mean that all the perturbative propagating modes that are present in the gravity sector at large $M$ are removed from the spectrum. These are the spin-2 ghost, the scalar tachyon and the massless graviton. The graviton in the confining phase now behaves like the confined gluon, a mass gap is generated for the graviton without a physical massive pole appearing in the propagator. It is this type of mass gap, which occurs in confining gauge theories, that we wish to highlight as a possibility for the graviton. But unlike QCD, the size of the mass gap here is controlled by a parameter $M$ that can be smaller than the strong scale $\Mp$. Then an interesting window opens up when $M\ll\Mp$. For $M^2\ll |k^2|\ll\Mp^2$ the propagator goes like $1/k^2$. The would-be perturbative behavior in this range is $-1/k^4$, but as before this is assumed to be reduced by the nonperturbative effect to $1/k^2$. And then as $k^2\to0$ the propagator tends to a negative constant instead of pole, just as for the gluon. For $|k^2|\gtrsim\Mp^2$ the behavior is $-1/k^4$. The resulting complete propagator can then be parametrized as $-G_M(k^2)/k^4$, with a new form factor $G_M(k^2)$ that is sensitive to $M$.

We illustrate $G_M(k^2)$ and $-G_M(k^2)/k^4$ by the red lines in Figs.~\ref{fig:GMpropagator}(a) and (b) respectively. For comparison, the blue line denotes the case at $M=0$. In the intermediate region $M^2\ll |k^2|\ll\Mp^2$, we see the approximation to $1/k^2$ behavior where the two color lines coincide. For smaller $|k^2|$ the deformation becomes significant and the would-be massless pole is removed due to $G_M(k^2)\sim k^4$ as $k^2\to0$ [as in Fig.~\ref{fig:Propagator}(b)]. Just like a high-energy gluon, the graviton propagates as a virtual particle but not as physical on-shell particle. For $|k^2|\gg M^2$ the difference with respect to $M=0$ can be negligible, as would be the case for most practical purposes if $M$ is of order the inverse of the size of the Universe. Finally in the limit $M=0$, the whole range below $\Mp$ has $1/k^2$ behavior and an on-shell massless graviton emerges. This limit appears to be smooth.

Before concluding this section we briefly consider the possibility that $M$ has some effect other than producing a graviton mass gap. The idea is that it is the massless spin-0 part of the metric field rather than the spin-2 part that receives a mass. Since the spin-0 mode is a constrained degree of freedom in general relativity, this would not change the propagating degrees of freedom. However this cannot be achieved in a local way. One possible covariant term in an effective action that affects the spin-0 mass is $R\, \square^{-2} R$, where $\square^{-1}$ is defined as an integration over a Green function. This particular term has been studied in the context of a phenomenological nonlocal modification of gravity in the IR~\cite{Maggiore}. There are issues of how causality is to be implemented.

We have proposed that the parameter $M$ controls the mass gap for the graviton and we have considered two different possibilities for the nature of this mass gap. A quantum phase transition as $M$ moves below $\Lambda_{\mathrm{QG}}\sim\Mp$ is of theoretical interest, but realistically $M$ needs to be very small or zero. In this regard we should emphasize the essential difference between our picture and the current development of massive graviton theories. For the latter, one of the key features is that diffeomorphism invariance is broken by a graviton mass and so there is a symmetry that is only restored in the massless limit. This then leads to the expectation that the graviton mass is stable under radiative corrections. In our picture the parameter $M$ that controls the graviton mass gap appears as a parameter in the underlying diffeomorphism-invariant theory. $M$ only breaks classical scale invariance, and so it can receive contributions in the high-energy regime, where the theory is perturbative, from any other explicit masses appearing in the matter sector. This is the price we have paid for our proposed UV-complete theory of a graviton mass gap.

\section{A focus on the measure}
\label{sec:refine}

If nonperturbative effects in quadratic gravity operate in a way similar to QCD we have argued that general relativity, or a modification of it that depends on $M$, can emerge in the IR. This is based on assuming that the nonperturbative multiplicative factors, $F(k^2)$ and $G(k^2)$, which we have introduced into the corresponding propagators, have a similar form. We are far from proving such a similarity, but we can explore some common nonperturbative features of the two theories. In this section we discuss how both QCD and quadratic gravity are based on path integrals over the space of gauge orbits, thus showing a fundamental similarity regarding the nontrivial measures and the nontrivial IR effects that are built in. In Sec.~\ref{sec:GCQCD} we discuss general properties of the measure due to the effects of Gribov copies and then relate this to the modification of the gluon propagator in QCD. In Sec.~\ref{sec:GCGravity} we search for Gribov copies for quadratic gravity. As a first step we show the existence of a Gribov horizon in the configuration space of gravity.

\subsection{Gribov copies in gauge theory}
\label{sec:GCQCD}

A gauge field configuration space has a redundancy, since a gauge transformation moves any gauge field configuration $A(x)$ along its gauge orbit. So the physical configuration space can be represented by the space of gauge orbits, which is the quotient space of the full configuration space modulo the group of local gauge transformations. Equivalently one can define a fundamental modular region (FMR) that intersects each gauge orbit only once. The path integral for a gauge theory can be defined by integrating over the FMR, but this definition proves not to be very practical. Faddeev and Popov (FP) proposed to insert the following factor of unity to extend the path integral to the full configuration space,
\begin{eqnarray}\label{eq:Identity}
1=\int \mathcal{D} U\, \delta(F(A^U))\det M_F(A)
.\end{eqnarray}
$F(A)=0$ is the gauge-fixing condition and $M_F(A)=\left.\delta F(A^U)/\delta U\right|_{F=0}$ is the FP operator.

However as first noticed by Gribov in 1977 \cite{Gribov:1977wm}, common gauge-fixing conditions do not properly restrict to the FMR. When considering finite-norm gauge configurations Gribov found that a gauge orbit could have a finite number of intersections with the gauge-fixing surface. In particular those gauge conditions that can be expressed as a variation of a norm functional, such as Landau and Coulomb gauge, are among those that are guaranteed to have Gribov copies according to the topological proof of Singer \cite{Singer:1978dk}. The number of intersections in excess of one is the number of Gribov copies $N_F(A)$. Once a gauge fixing $F(A)=0$ is chosen, $N_F(A)$ is gauge invariant, i.e.~it depends only on the orbit. $N_F(A)\neq0$ is only possible if the orbit probes the nonlinearities of the theory. On some such orbit there may be copies infinitesimally close to each other so that the gauge fixing becomes degenerate and the FP operator $M_F(A)$ has a zero eigenvalue. This defines the location of a Gribov horizon in configuration space.

In the presence of Gribov copies, we must replace the left-hand side of (\ref{eq:Identity}) by the total number of intersections, $1+N_F(A)$,
and the generating function is \cite{Gribov:1977wm}
\begin{eqnarray}\label{eq:ZFPGribov}
Z = \int \mathcal{D} A \,\frac{1}{1+N_F(A)}\delta (F(A))\left|\det M_F(A)\right|\,e^{i S(A)}
.\end{eqnarray}
The presence of the absolute value is dictated by the rules of calculus. For perturbative calculations there is no change from the FP prescription since $N_F(A)=0$ and $\det M_F>0$ for perturbative fields. In the nonperturbative regime, both $N_F(A)$ and the absolute value are important since the determinant can be negative on some copies. In practice there is no known way to completely absorb these effects into a modified action without approximation. As a result, the Gribov measure may prevent the construction of an exact BRST invariant gauge-fixed Lagrangian in the nonperturbative regime.

We note in passing that the naive FP prescription might still give the right generating function. As found in some models \cite{Friedberg:1995ty}\cite{Holdom:2009ws}, $N_F(A)$ is even and copies come in pairs with alternating sign for $\det M_F(A)$. In this case the signed intersection number is a topological invariant equal to unity for all gauge orbits~\cite{Hirschfeld:1978yq}. So if the absolute value is removed, all copies cancel out and $1/(1+N_F(A))$ should be replaced by unity. The trouble is that one ends up with a measure that alternates in sign and this is technically difficult to handle. Another way to tackle Gribov copies is to restrict the path integral to the region inside the first Gribov horizon, which is a region that includes the FMR. This is much easier to implement and it leads to the Gribov-Zwanziger action~\cite{Zwanziger:1989mf}\cite{RevGZ}. The first Gribov region also includes a set of copies that are by definition positive sign copies, $\det M_F>0$. The set of negative sign copies that are outside this region and that would cancel these positive sign copies are omitted from the path integral. Thus one is left to wonder just how accurate such an approximation can be.

The Gribov measure in (\ref{eq:ZFPGribov}) captures effects that are built into the nonperturbative definition of the theory, and thus it provides a direct view into nonlocal and topologically interesting features of the theory. By inspection of the Gribov equation, which is the condition that copies must satisfy, $N_F(A)$ is scale invariant. That is two orbits related by a scale transformation have the same number of copies. So in a scale-invariant gauge theory, for example one that is at a fixed point, Gribov copies would have nontrivial effects on all scales in a scale-invariant way.\footnote{An approximation that introduces a mass scale would fail to describe this. Proposals to deal with Gribov copies usually introduce a mass scale into some modified action. This is a failure of the approximation given that Gribov copies are intrinsically scale invariant.} On the other hand in an asymptotically free gauge theory the scale $\Lambda$ at which the coupling grows strong will be the scale at which Gribov copies are important.

There are also implications of the fact that $N_F(A)$ is an integer. In particular $N_F(A)\equiv0$ for a certain bounded region of gauge configurations within the FMR. This region includes the perturbative regime. Consider a configuration that can be characterized by a typical momenta $k$ and an amplitude $A_k$. Since $N_F(A)$ is scale invariant, $N_F(A)$ should depend on the scale-invariant ratio $A_k/k$. From this we expect that $N_F(A)$ becomes nonzero at some critical value of $A_k$ proportional to $k$. This may be checked explicitly from the Gribov equation \cite{Holdom:2009ws}. By dimensional analysis then Gribov copies are only important when $A_k^2 \gtrsim {k^2}/{\Lambda^4}$.
On the other hand, a gauge field fluctuation has typical size $A_k^2 \approx {1}/{k^2}$ according to the propagator. Thus at large $k^2\gg\Lambda^2$ the size of typical vacuum fluctuations are much smaller than needed to feel any effect of Gribov copies.

Since the path integral involves $\sim\exp( -\sum_k k^2 A_k^2)$, the fluctuations that are large enough to feel Gribov copies are exponentially suppressed by $\sim \exp(-k^4/\Lambda^4)$. From this we conclude that the corrections to the propagator due to Gribov copies at high $k^2$ are exponentially small. Earlier we introduced the $F(k^2)$ into the gluon propagator to model nonperturbative effects. If we use this to model the effects of the measure due to Gribov copies, then the deviation of $F(k^2)$ from unity at large $|k^2|$ should be $\sim \exp(-k^4/\Lambda^4)$. Thus the extreme nonlocality of Gribov copies makes their effects extremely soft in the UV.\footnote{This is to be contrasted with the power-law corrections that are present for example in the Gribov-Zwanziger approach. There the propagator has a pair of complex-conjugate poles on the complex $k^2$ plane, and this would imply power-law corrections in the UV that are excluded by our argument.}

However in the IR, the path integral easily samples configurations with amplitudes above the critical value, at which point $N_F(A)$ can grow very quickly. This implies a suppressed propagator in the IR, with the explicit behavior in the $k^2\to0$ limit depending in detail on how $N_F(A)$ grows in the large fluctuation region. For Coulomb gauge and spherically symmetric configurations, a certain power-law growth was found as a function of $A_k/k$~\cite{Holdom:2009ws}. This implied a vanishing gluon propagator in the $k^2\to0$ limit with $F(k^2)\sim k^4$, for example as shown in Fig.~\ref{fig:Propagator}(b).

\subsection{Gribov copies in gravity}
\label{sec:GCGravity}

Gravitation reflects the dynamics of the spacetime continuum and it is described by a theory invariant under coordinate transformations. Taking the metric $g_{\mu\nu}$ to be the fundamental dynamical field, under coordinate transformations it transforms as
\begin{eqnarray}
g_{\mu\nu}(x)=\frac{\partial x'^{\alpha}}{\partial x^{\mu}}\frac{\partial x'^{\beta}}{\partial x^{\nu}}g'_{\alpha\beta}(x')
.\end{eqnarray}
The general covariance can also be described as a gauge transformation of the metric field with pullback diffeomorphism \cite{Wald:1986bj}. The full metric is split as $g_{\mu\nu}=\bar{g}_{\mu\nu}+h_{\mu\nu}$ and $h_{\mu\nu}$ is treated as a dynamical symmetric tensor field on the manifold equipped with metric $\bar{g}_{\mu\nu}$. For an infinitesimal coordinate transformation $x'^\mu=x^\mu+\xi^\mu$, the corresponding gauge transformation of $h_{\mu\nu}$ is the Lie derivative along $\xi^\mu$
\begin{eqnarray}\label{eq:GravityGauge}
\delta_\xi h_{\mu\nu}=\mathcal{L}_\xi g_{\mu\nu}
=\bar{\nabla}_\nu \xi_\mu+\bar{\nabla}_\mu \xi_\nu
+\xi^\rho\bar{\nabla}_\rho h_{\mu\nu}+h_{\mu\rho}\bar{\nabla}_\nu \xi^\rho+h_{\rho\nu}\bar{\nabla}_\mu \xi^\rho
,\end{eqnarray}
where $\bar{\nabla}_\mu$ is the covariant derivative with respect to $\bar{g}_{\mu\nu}$ and $\xi_\mu=\bar{g}_{\mu\nu}\xi^\nu$.

With a background $\bar{g}_{\mu\nu}$ and a gauge-fixing condition $F(h_{\mu\nu})=0$, the quantum theory of gravity can be defined by the path integral in analogy to a gauge theory in a background field gauge,
\begin{eqnarray}\label{eq:ZFPGribovG}
Z = \int \mathcal{D} h \,\frac{1}{1+N_F(h)}\delta (F(h))\left|\det M_F(h)\right|\,e^{i  S_{\mathrm{QG}}(g)}
.\end{eqnarray}
$M_F(h)=\left.\delta F(h^\xi)/\delta \xi\right|_{F=0}$ is the FP operator and $\xi$ denotes the gauge transformation in (\ref{eq:GravityGauge}). This construction of the measure only involves the symmetry of the theory rather than the explicit form of the action. Since quadratic gravity is asymptotically free, if Gribov copies exist they will produce a nonperturbative effect in the IR through the presence of $N_F(h)$ in the measure. The Gribov problem in the gravity has been rather sparsely studied~\cite{GribovGravity1}\cite{GribovGravity2}. As a first step along this line we will examine the Gribov horizon equation, the infinitesimal version of the Gribov equation, with respect to some background $\bar{g}_{\mu\nu}$. This requires a solution to $F(\delta_\xi h_{\mu\nu})=0$ for some $h_{\mu\nu}$ and some $\xi^\mu$. This is the same as finding a zero eigenvalue of the FP operator.

First we  consider de Donder gauge where $F(h)=\bar{\nabla}^\mu h_{\mu\nu}-\frac{1}{2}\bar{\nabla}_\nu h=0$. The flat background is the first case to consider, and the Gribov horizon equation in this gauge is as follows,
\begin{eqnarray}
\partial^\mu\partial_\mu \xi_\nu
+\partial^\mu \xi^\rho \partial_\rho h_{\mu\nu}-\frac{1}{2}\partial_\nu\xi^\rho\partial_\rho h
+\partial^\mu h_{\mu\rho}\partial_\nu \xi^\rho-\partial_\nu h_{\mu\rho}\partial^\mu \xi^\rho
+\partial^\mu(h_{\rho\nu}\partial_\mu\xi^\rho)=0
\label{simpGH}
.\end{eqnarray}
As expected there are no copies when $h_{\mu\nu}=0$. In analogy to Gribov's approach to QCD, we study static spherically symmetric metrics and require that $h_{\mu\nu}$ has a finite norm
\begin{align}
{\cal N}=\int d^3x \sqrt{-\bar g} h_{\mu\nu}h^{\mu\nu}
,\end{align}
calculated with respect to the background metric. It turns out to be difficult to find a nontrivial solution of (\ref{simpGH}) such that $h_{\mu\nu}$ has finite norm and $\xi_\mu$ is well behaved. The equation (\ref{simpGH}) is linear in $\xi_\mu$, but since $\xi_\mu$ represents an infinitesimal transformation it should be a bounded function.

We now note that in de Donder gauge, $F(h)=0$ is not obtained as the vanishing gauge variation of the norm function. Thus in this sense it is not the analog of the Landau gauge or covariant background field gauges in QCD, where the gauge-fixing plane is a collection of stationary points of the norm. In gravity the variation of the norm function $\cal N$ turns out to be nonlinear in $h_{\mu\nu}$. Requiring that this variation vanishes defines a nonlinear gauge-fixing condition for gravity, which we could call the norm gauge. In the following we derive the Gribov horizon equations in norm gauge and in this case we have more success in solving these equations.

We consider a flat background and a static choice for $h_{\mu\nu}$. The line element is $ds^2=ds^2_{\bar g}+ds^2_h$ where
\begin{align}
ds^2_{\bar g}&=-dt^2+dr^2+r^2d\theta^2+r^2\sin(\theta)^2d\phi^2,\nonumber\\
ds^2_h&=i(r)dt^2+2k(r)dtdr+h(r)dr^2
.\end{align}
$ds^2_h$ retains the same form under the infinitesimal transformation generated by $\xi^\mu = (\alpha(r),\beta(r),0,0)$, where
{\small\begin{align}
\delta i(r) &=\beta(r) i'(r),\nonumber\\
\delta k(r) &= \beta'(r) k(r) + \alpha'(r)  \left( i(r) -1 \right)+\beta(r) k'(r),\nonumber\\
\delta h(r) &=2\, \beta'(r)  \left( h(r) +1 \right) +2\, \alpha'(r) k(r) +\beta(r) h'(r).
\label{e1}\end{align}}
The norm function is
\begin{align}
{\cal N}=4\pi \int dr r^2(h(r)^2+i(r)^2-2k(r)^2)
.\label{e4}\end{align}
If we require that the norm function vanishes under the $\alpha(r)$ and $\beta(r)$ variations separately, this gives two nonlinear gauge-fixing conditions.
{\small\begin{align}
&k'(r)\;[ r+r h(r)- r i(r)]+k(r)\;[2+2\,h(r)  -2\,i(r)+r h'(r)- r i'(r) ]=0\label{e2}\\
&4\,h(r)+4\, h(r)^2-4\, k(r)^2+2\,r h'(r)+3\,r h(r) h'(r)- r i(r) i'(r)-2\,r k(r) k'(r)=0\label{e3}
\end{align}}
The first can be solved trivially with $k(r)=0$, and we will adopt this in the following.

Making a further infinitesimal transformation on (\ref{e2}) using (\ref{e1}) gives the first Gribov horizon equation. This equation does not depend on $\beta(r)$ and it can be reduced to
{\small\begin{align}
\alpha'(r) =\frac {1}{r^2 [1+h(r) -2\,i(r)+ i(r)^2-h(r) i(r)]}
.\label{e5}\end{align}}
The case $i(r)=0$ and $h(r)=0$ would imply that $\alpha(r)$ is not bounded and so cannot represent an infinitesimal transformation. A strategy then is to choose a trial $i(r)$ and then use the second gauge-fixing condition (\ref{e3}) to solve for $h(r)$. The latter must be done numerically. The goal is to find $i(r)$ and $h(r)$ having a finite norm $\cal N$ in (\ref{e4}) and that result in a bounded $\alpha(r)$ from (\ref{e5}). In fact we find that there is a whole family of solutions of this type, with $i(r)=-1/(r+r^2)$ being one example. Note that such configurations are of finite norm even though they are singular at $r=0$.

The second Gribov horizon equation can be similarly derived from the second gauge-fixing condition (\ref{e3}). It only depends on $\beta(r)$,
{\small\begin{align}
&\beta''(r)\;[4\,r+10\,r h(r)+6\,r h(r)^2 ] \nonumber\\
&+\beta'(r)\;[8+24\,  h(r)+16\,h(r)^2+12\,r  h'(r)+15\,rh(r)   h'(r) -r  i(r)  i'(r)  ]  \nonumber\\
&+\beta(r)\;[4\, h'(r)+8\,h(r)  h'(r)-r  i'(r)^2+3\,r  h'(r)^2 +2\, r h''(r)+3\,r h(r) h''(r)- r i(r) i''(r)]=0
.\label{e6}\end{align}}
A suitable solution to this equation would give a second Gribov horizon independent of the first. The two horizons would correspond to two gauge-fixing degeneracies with respect to different gauge transformations and there is no requirement that these horizons exist for the same $h_{\mu\nu}$. But in fact we find that they do; there are $i(r)$ and $h(r)$ that we found from the first horizon equation that also yield a numerical solution for $\beta(r)$ from (\ref{e6}) that is bounded. [A slight difference in this case is that $\beta'(r)$ is not bounded.]

Thus we have found finite-norm solutions in norm gauge that satisfy one or both of the horizon equations. This indicates that the gauge-fixing condition has become degenerate, or in other words that a pair of Gribov copies are connected by some infinitesimal coordinate transformation $\xi^\mu$. Then we would also expect pairs of copies on either side of the horizon that are connected by finite transformations. Thus we find evidence for Gribov copies in gravity in a covariant gauge based on a norm function just as in QCD. And by the same analogy we can expect that the number of copies $N_F(h)$ is finite for finite-norm configurations.

There is another similarity with the gauge theory, namely that $N_F(h)$ is scale invariant. A scale transformation can be implemented as a combined Weyl rescaling of the metric and a coordinate transformation such that $ds^2\to \lambda^2ds^2$ and $h_{\mu\nu}(x)\to h_{\mu\nu}(\lambda x)$. The Lie derivative (\ref{eq:GravityGauge}) transforms homogeneously as does the gauge-fixing condition $F(h)=0$. In norm gauge the latter is due to the norm itself being homogeneous in $h_{\mu\nu}$. Thus the Gribov horizon equation also transforms homogeneously, as seen for example in (\ref{e6}), which will thus have solutions related by scale transformations. Similarly the full Gribov equation will have the same property. Thus two gauge orbits related by a scale transformation have the same number of copies, meaning that $N_F(h)$ is scale invariant.

Interestingly there is a family of background metrics with naked timelike singularities, including the negative-mass Schwartzchild metric and the nakedly singular Reissner-Nordstrom metric, where a Gribov horizon exists even for vanishing $h_{\mu\nu}$, in both de Donder gauge and norm gauge. This means the gauge-fixing fails even in the perturbative regime upon these backgrounds. The fact that a perturbative quantum theory of gravity cannot be defined on such nakedly singular backgrounds may be an interesting result on its own. We shall return to this in the next section.

The existence of Gribov horizons for certain spherically symmetric and static metrics in a covariant gauge provides us with some evidence for the fundamental similarity between quadratic gravity and QCD. In gravity much less is known about the nature of the FMR or the shape of Gribov horizons, but the path integral should again be affected by the $1/(1+N_F(h))$ factor in the measure. The basic properties of this factor that we discussed for asymptotically free gauge theories should continue to hold here. Thus we expect that the Gribov measure does not produce power-law corrections in the UV and in particular that the perturbative propagator is approached exponentially quickly in the UV. This means that the nonperturbative physics that we are discussing is not associated with vacuum condensates or complex-conjugate poles in propagators or anything else that would imply power-law corrections.

In the IR the Gribov measure effectively reduces the size of the metric field configuration space, which is manifested as a suppressed propagator. We have used the similarity to QCD to suggest a propagator suppression by a factor of $-k^2$ for $|k^2|\ll\Lambda_{\mathrm{QG}}^2$ and $M=0$. Studies that could shed more light on the nature of the space of gauge orbits, such as solutions of the full Gribov equation, could further test the similarity with QCD. Perhaps most decisive would be a lattice formulation of an asymptotically free theory of gravity in complete analogy to lattice studies of QCD.

So far all our discussions are based on the original metric field, with additional degrees of freedom hidden in higher-derivative propagators. A common practice is to make explicit these degrees of freedom via the auxiliary field method. With the introduction of a symmetric tensor field $f_{\mu\nu}$, the Weyl tensor term can be replaced by a kinetic mixing term $f_{\mu\nu}G^{\mu\nu}$ ($G^{\mu\nu}$ is the Einstein tensor) and a mass term $(f_{\mu\nu}f^{\mu\nu}-f^2)$. Expanding around the flat background, the quadratic action has a kinetic mixing between $h_{\mu\nu}$ and $f_{\mu\nu}$. When $M$ is large, the diagonal basis gives rise to the expected particle content in the perturbative phase, i.e. the normal graviton and massive spin-2 ghost. With decreasing $M$ the kinetic mixing becomes more important. When $M\lesssim \Mp$, this implies not only strong dynamics for both fields but also a strong interaction between the two sectors. As we discussed in this section, the nontrivial measure in the path integral plays a crucial role for the nonperturbative physics in the IR. Given the strong mixing, the two-fields language becomes less convenient to conduct an analogy with QCD for the nonperturbative modification of the propagators.

Finally we give two examples for the nonperturbative multiplicative factor $F(k^2)$ as displayed in Fig.~\ref{fig:Propagator},
\begin{eqnarray}
(\textrm{a})\,\,\, \frac{k^2}{k^2+b/(k^2-a)},\quad
(\textrm{b})\,\,\,\frac{{\rm erf}(a)-{\rm erf}(b k^2)-{\rm erf}(-b k^2+a)}{{\rm erf}(a)}
.\end{eqnarray}
${\rm erf}(x)$ is the error function. $a, b>0$ in each case gives the qualitative shape in Fig.~\ref{fig:Propagator}(a) while the $a\rightarrow0$ limit in each case gives Fig.~\ref{fig:Propagator}(b). Case (a) with $a, b>0$ corresponds to the refined Gribov-Zwanziger propagator with complex-conjugate poles in the complex $k^2$ plane~\cite{RefineGZ}. Complex-conjugate poles also appear in the perturbative modification of the graviton propagator as proposed in~\cite{Tomboulis:1977jk}. Case (b) instead has exponentially small effects in the UV as we expect from Gribov copies. In fact the deviation from unity is suppressed by $\exp(-b^2k^4)$ for large $|k^2|$ as we discussed above. It is also an entire function, which brings in some connection with the recent study of nonlocal field theory in~\cite{Tomboulis:2015gfa}. Nonlocal propagators in a perturbative context would imply associated acausal effects, but the nonlocality in our propagators is occurring where there is no perturbative description. In the case of gravity this is near $\Mp$, and in the low-energy effective theory these nonlocal effects are encoded in the higher-derivative terms in the derivative expansion.

\section{Implications and speculations}
\label{sec:implication}

We first consider the quantum effects that an asymptotically free theory of gravity can have on the matter sector. In our picture these effects are quite minimal, since there is only a limited range of energies close to the Planck mass where the gravitational interactions are strong. This can produce order-one multiplicative effects on the running of the matter couplings as they run through this region, but outside this region the gravity contributions are perturbative. Thus a normal QFT description of the matter sector persists on super-Planckian energy scales. Any problems with the matter sector, such as the existence of UV Landau poles for the $U(1)$ hypercharge coupling or the Higgs quadratic coupling, should be solved within the matter sector. To insist that all couplings be asymptotically free in a theory with elementary scalar fields turns out to be a very nontrivial requirement. Extensions of the standard model where the ratios of all couplings, including scalar couplings, approach a UV fixed point have been studied, both for UV unstable~\cite{SAFE1} and UV  stable~\cite{SAFE2} cases. Non-Gaussian fixed points (UV unstable) have also been considered~\cite{Litim:2014uca}.

For the rest of this section we consider the more direct gravitational implications of asymptotically free quadratic gravity when $M=0$. As curvatures increase the higher-order terms in a derivative expansion become more important, as in the standard picture, but at sufficiently high curvatures the theory simplifies and higher-derivative terms past four are absent. In the far UV the couplings are arbitrarily small and then there are only small quantum gravity fluctuations. The theory could be fundamentally defined around flat spacetime, for example on a lattice with sufficiently small lattice spacing. To incorporate the quantum gravity corrections around some other background, an effective action with the small scale corrections integrated out down to the typical curvature scale of this other background should be used. If this scale is still well above $\Mp$, then we can expect these quantum corrections to be small, and so the classical description could be a good approximation. If this type of picture from asymptotically free gauge theories is correct for gravity then it leads one to take more seriously classical solutions of quadratic gravity in regions of curvature above $\Mp$, for example to address the singularity problems of general relativity.

The classical analysis of the action in (\ref{eq:quadratic}) is usually done when $M$ is identified with $\Mp$, and for spherically symmetric solutions it yields three classes of solutions at $r=0$~\cite{QGBH1}. One class is simply the isolated Schwarzschild solution while another class is nonsingular at $r=0$.  The third class is singular, but in~\cite{QGBH2} it was found that the number of parameters that characterize this class of solutions at $r=0$ matches the number that are expected at large $r$ in a four-derivative theory. This then leads to a rather generic set of solutions that only deviate from the Schwarzschild solution by exponentially small terms at large $r$. At smaller $r$ these solutions have no horizon, but instead large curvatures turn on close to where the horizon would have been. The singularity at $r=0$ in this vacuum solution is timelike, and a complete solution of this type was obtained numerically in~\cite{QGBH2}. Such solutions were studied further in~\cite{QGBH3}.

Our present picture (with $M=0$) is somewhat different. In the small-curvature region we expect that the classical solutions of interest are determined by the derivative expansion theory in (\ref{eq:derexp}), where the Einstein-Hilbert term dominates. For super-Planckian curvatures it is the fundamental pure quadratic action that controls the solutions. In these two limits we have suggested that the quantum corrections are small and then the solutions should resemble those of the previous studies. But for Planck-sized curvatures the quantum corrections are expected to be large. Thus a strongly interacting shell should appear at some radius in the spherically symmetric solutions, that otherwise could be similar to the previous analyses. For the Schwarzschild-like solution, this shell should appear deep inside the interior of the black hole. For the horizonless solutions, the strong shell should appear close to where the horizon would have been. For smaller $r$ the curvature of the vacuum solution increases and the description should become more classical. In the purely classical analysis there was no reason to stop at a four-derivative description of the interior region, so our present picture puts these strong horizonless objects on a firmer footing.\footnote{The timelike singularity in this case does not suffer from the perturbative gauge-fixing problem mentioned in the previous section.}

When we consider the gravitational field around any spherical mass, the derivative expansion (\ref{eq:derexp}) describes the large-$r$ region. We can expect that this will generate corrections to the gravitational potential, at the very least by terms that are exponentially small. This is due to the effectively massive modes, in particular modes with positive $m^2\sim\Mp^2$, that the derivative expansion implies. Thus Yukawa-type corrections appear to be quite generic, but it is typical to assume that they are always insignificant. The surprise from the horizonless solutions in the classical analysis is that the Yukawa terms can become significant just where the standard gravitational potential approaches order one. When the massive modes turn on, the curvatures quickly become large, the derivative expansion breaks down and the strong gravity regime is entered. It seems to us that the effect of massive modes should be taken into account in the description of gravitational collapse, and that this will have a bearing on deciding what actually does form, black holes or strong horizonless objects.

As for other implications of a UV-complete theory, one might wonder whether a super-Planckian scattering experiment could in principle probe the region where the hard scattering process is under the theoretical control of an asymptotically free theory. This is the obvious analog of $pp$ scattering at the LHC, where perturbative QCD calculations are instrumental for standard model predictions. The analogs of factorization, parton distribution functions, parton showers, and hadronization could seemingly all be carried over to the gravity description. In analogy with perturbative QCD, to the extent that a hard process can be defined to be independent of the strong IR physics, that hard process involves virtual gravitons that are far off shell (with $1/k^4$ behavior) and which are thus insensitive to the strong IR physics that has altered their on-shell behaviors.

But we run into a major difference for super-Planckian scattering in gravity, and that is the formation of black holes. In fact a detailed picture is already emerging~\cite{Eardley:2002re} where it is purported that when $\sqrt{s}\gg\Mp$, semiclassical black holes of mass around $\sqrt{s}$ dominate the production cross section. At least under some conditions a description of the formation of closed trapped surfaces is given without needing knowledge of the super-Planckian theory~\cite{Giddings:2004xy}. Considering also the possible loss of unitarity due to the black hole information problem, this leads to a completely different and still incomplete picture compared to the manifestly unitary, jet-like final states from high-energy scattering in QCD. In view of this conundrum it could be of interest to consider the production of strong horizonless objects in super-Planckian scattering as well.

We turn to another distinctive aspect of our theory. This is the fact that upon increasing the parameter $M$ sufficiently, there is a phase transition to a theory that is manifestly unstable. This leads to the question of whether the $M=0$ theory could be pushed into instability somehow within the theory itself, in particular for example in a region of sufficiently large curvature. In principle the metric fluctuations around different background metrics could have an altered spectrum, and for backgrounds with super-Planckian curvatures it could be possible that the spectrum picks up unstable modes. One possibility is that the curvature acts like an IR cutoff, effectively eliminating the strong interaction effects that removed the ghost in the first place. An example of an IR cutoff that could do this occurs for a compact universe with size much smaller than $1/\Mp$. This would likely exhibit an instability. A universe at temperature $T\gg\Mp$ might as well.

In our study of Gribov copies, we found that Gribov horizons did not tend to disappear in the case of asymptotically flat background metrics with localized high curvatures. In fact for certain metrics the effects of Gribov copies became even more pronounced by afflicting the perturbative fluctuations; that is the Gribov horizon existed for vanishing $h_{\mu\nu}$. So in this case the curvature does not act like an IR cutoff to remove the nonperturbative physics. But the fact that the perturbative quantum description is itself breaking down around such backgrounds (at least for the gauges we considered) could itself be a sign that the spectrum has changed. The backgrounds in question have naked, timelike singularities.

We thus entertain the possibility that unstable modes could develop under conditions of high curvatures of a type that are approaching timelike singularities. Such conditions may be quite generic during the final stages of gravitation collapse~\cite{Joshi:2013xoa}, and they could occur in the super-Planckian scattering example. If an unstable mode develops within the region of high curvature then the analog of vacuum decay can occur, where positive- and negative-energy modes are spontaneously created with arbitrarily high momenta while conserving total energy. The interesting point is that the negative modes are confined to propagate within the region where they can exist, whereas the positive modes can escape and ``hadronize'' into normal gravitons and other particles. The negative modes will instead interact with positive-energy matter and thus lower the energy of this normal matter in the high-curvature region. The net result could be an explosive release of energy that should reduce the curvature and take the region away from the quantum instability. Such a burst of energy in gravitational collapse may be of interest as a source for ultra high-energy cosmic rays~\cite{Torres:2004hk}.

\begin{acknowledgments}

This research is supported in part by the Natural Sciences and Engineering Research
Council of Canada. J.R. is also supported in part by the International Postdoctoral Exchange
Fellowship Program of China. We thank Chen Zhang for early collaboration and related discussions. We are grateful for useful discussions with Niayesh Afshordi, Asimina Arvanitaki, Cliff Burgess, John Donoghue, Kurt Hinterbichler, Luis Lehner, Sakura Schafer-Nameki, Eva Silverstein, Yong Tang, John Terning, Jesse Thaler, Terry Tomboulis and Zhong-Zhi Xianyu.

\end{acknowledgments}	

\linespread{1}

\begin{thebibliography}{99}

\bibitem{Weyl}
  H.~Weyl,
  Ann.~Phys.\  {\bf 59}, 101 (1919); Surveys High Energ.\ Phys.\  {\bf 5}, 237 (1986).

\bibitem{Donoghue:1994dn}
  J.~F.~Donoghue,
  Phys.\ Rev.\ D {\bf 50}, 3874 (1994);
  AIP Conf.\ Proc.\  {\bf 1483}, 73 (2012);
  J.~F.~Donoghue and B.~R.~Holstein,
  J.\ Phys.\ G {\bf 42}, 103102 (2015).

\bibitem{Burgess:2003jk}
  C.~P.~Burgess,
  Living Rev.\ Relativity\  {\bf 7}, 5 (2004).

\bibitem{Stelle:1976gc}
  K.~S.~Stelle,
  Phys.\ Rev.\ D {\bf 16}, 953 (1977).

\bibitem{Voronov:1984kq}
  B.~L.~Voronov and I.~V.~Tyutin,
  Yad.\ Fiz.\  {\bf 39}, 998 (1984).

\bibitem{GBterm1}
  G.~de Berredo-Peixoto and I.~L.~Shapiro,
  Phys.\ Rev.\ D {\bf 71}, 064005 (2005).

\bibitem{GBterm2}
  M.~B.~Einhorn and D.~R.~T.~Jones,
  Phys.\ Rev.\ D {\bf 91}, 084039 (2015).

\bibitem{Fradkin:1981iu}
  E.~S.~Fradkin and A.~A.~Tseytlin,
  Nucl.\ Phys.\ {\bf B201}, 469 (1982).

\bibitem{Avramidi:1985ki}
  I.~G.~Avramidi and A.~O.~Barvinsky,
  Phys.\ Lett.\ B {\bf 159}, 269 (1985).

\bibitem{Agravity1}
  A.~Salvio and A.~Strumia,
  JHEP {\bf 06} (2014) 080.

\bibitem{Agravity2}
  M.~B.~Einhorn and D.~R.~T.~Jones,
  JHEP {\bf 03} (2015) 047.

\bibitem{ADM}
  R.~L.~Arnowitt, S.~Deser and C.~W.~Misner,
  Gen.\ Rel.\ Grav.\  {\bf 40}, 1997 (2008).

\bibitem{Ostrogradski}
  M. Ostrogradski, Mem. Ac. St. Petersbourg VI, 385 (1850).

 \bibitem{Woodard:2015zca}
  R.~P.~Woodard, Scholarpedia {\bf 10}, 32243 (2015).

\bibitem{WW}
  S.~Weinberg and E.~Witten,
  Phys.\ Lett.\ B {\bf 96}, 59 (1980).

\bibitem{Gribov:1977wm}
  V.~N.~Gribov,
  Nucl.\ Phys.\ {\bf B139}, 1 (1978).

\bibitem{analogy}
  C.~T.~Hill,
 arXiv:hep-ph/0510177.

\bibitem{Maggiore:2015rma}
  M.~Maggiore,
  arXiv:1506.06217.

\bibitem{Tomboulis:1977jk}
  E.~Tomboulis,
  Phys.\ Lett.\ B {\bf 70}, 361 (1977);
  Phys.\ Lett.\ B {\bf 97}, 77 (1980).
  
\bibitem{leewick}
T.~D.~Lee and G.~C.~Wick,
  Nucl.\ Phys.\ {\bf B9}, 209 (1969);
  Nucl.\ Phys.\ {\bf B10}, 1 (1969);
  Phys.\ Rev.\ D {\bf  2}, 1033 (1970).

\bibitem{Antoniadis:1986tu}
  I.~Antoniadis and E.~T.~Tomboulis,
  Phys.\ Rev.\ D {\bf 33}, 2756 (1986).

\bibitem{Johnston}
  D.~A.~Johnston,
  Nucl.\ Phys.\ {\bf B297}, 721 (1988).

\bibitem{Pais:1950za}
  A.~Pais and G.~E.~Uhlenbeck,
  Phys.\ Rev.\  {\bf 79}, 145 (1950).

\bibitem{ToyModel1}
  A.~van Tonder,
  arXiv:0810.1928.

\bibitem{ToyModel2}
  M.~Pavsic,
  Mod.\ Phys.\ Lett.\ A {\bf 28}, 1350165 (2013).

\bibitem{Chen:2013aha}
  T.~j.~Chen and E.~A.~Lim,
  JCAP {\bf 05} (2014) 010.

\bibitem{Mannheim}
  P.~D.~Mannheim,
  Found.\ Phys.\  {\bf 37}, 532 (2007);
  C.~M.~Bender and P.~D.~Mannheim,
  Phys.\ Rev.\ Lett.\  {\bf 100}, 110402 (2008).

\bibitem{Salvio:2015gsi}
  A.~Salvio and A.~Strumia,
  Eur.~Phys.~J.~ C {\bf 76}, 227 (2016).

\bibitem{Horava:2009uw}
  P.~Horava,
  Phys.\ Rev.\ D {\bf 79}, 084008 (2009).

\bibitem{Tomboulis1997}
  E.~T.~Tomboulis,
  arXiv:hep-th/9702146.
  
\bibitem{Modesto:2011kw} 
 L.~Modesto,
 Phys.\ Rev.\ D {\bf 86}, 044005 (2012).

\bibitem{Biswas:2011ar}
  T.~Biswas, E.~Gerwick, T.~Koivisto and A.~Mazumdar,
  Phys.\ Rev.\ Lett.\  {\bf 108}, 031101 (2012).

\bibitem{RGE1}
  A.~Salam and J.~A.~Strathdee,
  Phys.\ Rev.\ D {\bf 18}, 4480 (1978).
  
\bibitem{RGE3}
  G.~Narain and R.~Anishetty,
  Phys.\ Lett.\ B {\bf 711}, 128 (2012);
  G.~Narain and R.~Anishetty,
  J.\ Phys.\ Conf.\ Ser.\  {\bf 405}, 012024 (2012).

\bibitem{RGE2}
  D.~Benedetti, P.~F.~Machado and F.~Saueressig,
  Mod.\ Phys.\ Lett.\ A {\bf 24}, 2233 (2009).

\bibitem{Tomboulis:1983sw}
  E.~T.~Tomboulis,
  Phys.\ Rev.\ Lett.\  {\bf 52}, 1173 (1984).

\bibitem{Alkofer:2000wg}
  For a review of QCD Green functions, see R.~Alkofer and L.~von Smekal,
  Phys.\ Rep.\  {\bf 353}, 281 (2001).

\bibitem{Oehme:1979ai}
  R.~Oehme and W.~Zimmermann,
  Phys.\ Rev.\ D {\bf 21}, 471 (1980).

\bibitem{Cornwall:2013zra}
  J.~M.~Cornwall,
  Mod.\ Phys.\ Lett.\ A {\bf 28}, 1330035 (2013).

\bibitem{SDS}
  L.~von Smekal, R.~Alkofer and A.~Hauck,
  Phys.\ Rev.\ Lett.\  {\bf 79}, 3591 (1997).

\bibitem{SDD}
  A.~C.~Aguilar, D.~Binosi and J.~Papavassiliou,
  Phys.\ Rev.\ D {\bf 78}, 025010 (2008).

\bibitem{Zwanziger:1989mf}
  D.~Zwanziger,
  Nucl.\ Phys.\ {\bf B323}, 513 (1989).
  
\bibitem{RevGZ}
  N.~Vandersickel and D.~Zwanziger,
  Phys.\ Rep.\  {\bf 520}, 175 (2012).

\bibitem{RefineGZ}
  D.~Dudal, J.~A.~Gracey, S.~P.~Sorella, N.~Vandersickel and H.~Verschelde,
  Phys.\ Rev.\ D {\bf 78}, 065047 (2008).

\bibitem{LGlattice1}
  A.~Cucchieri and T.~Mendes,
  Proc.~Sci.~LAT (2007) 297,
 arXiv:0710.0412.

\bibitem{LGlattice2}
  I.~L.~Bogolubsky, E.~M.~Ilgenfritz, M.~Muller-Preussker and A.~Sternbeck,
  Proc.~Sci.~LAT (2007) 290,
  arXiv:0710.1968.

\bibitem{Nakagawa:2009zf}
  Y.~Nakagawa, A.~Voigt, E.-M.~Ilgenfritz, M.~Muller-Preussker, A.~Nakamura, T.~Saito, A.~Sternbeck and H.~Toki,
  Phys.\ Rev.\ D {\bf 79}, 114504 (2009).

\bibitem{RevMassG1}
  K.~Hinterbichler,
  Rev.\ Mod.\ Phys.\  {\bf 84}, 671 (2012).

\bibitem{RevMassG2}
  C.~de Rham,
  Living Rev.\ Relativity  {\bf 17}, 7 (2014).

\bibitem{Cline:2003gs}
  J.~M.~Cline, S.~Jeon and G.~D.~Moore,
  Phys.\ Rev.\ D {\bf 70}, 043543 (2004).

\bibitem{Maggiore}
  M.~Maggiore and M.~Mancarella,
  Phys.\ Rev.\ D {\bf 90}, 023005 (2014).

\bibitem{Singer:1978dk}
  I.~M.~Singer,
  Commun.\ Math.\ Phys.\  {\bf 60}, 7 (1978).

\bibitem{Friedberg:1995ty}
  R.~Friedberg, T.~D.~Lee, Y.~Pang and H.~C.~Ren,
  Ann.~Phys.\  {\bf 246}, 381 (1996).

\bibitem{Holdom:2009ws}
  B.~Holdom,
  Phys.\ Rev.\ D {\bf 79}, 085013 (2009).

\bibitem{Hirschfeld:1978yq}
  P.~Hirschfeld,
  Nucl.\ Phys.\ {\bf B157}, 37 (1979).

\bibitem{Wald:1986bj}
  R.~M.~Wald,
  Phys.\ Rev.\ D {\bf 33}, 3613 (1986).

\bibitem{GribovGravity1}
  A.~K.~Das and M.~Kaku,
  Nuovo Cimento Soc.~Ital.~Fis.~{\bf 50B}, 303 (1979).

\bibitem{GribovGravity2}
  A.~Anabalon, F.~Canfora, A.~Giacomini and J.~Oliva,
  Phys.\ Rev.\ D {\bf 83}, 064023 (2011).

\bibitem{Tomboulis:2015gfa}
  E.~T.~Tomboulis,
Phys.~Rev.~D {\bf 92}, 125037 (2015).

\bibitem{SAFE1}
  G.~F.~Giudice, G.~Isidori, A.~Salvio and A.~Strumia,
  JHEP {\bf 02} (2015) 137.

\bibitem{SAFE2}
  B.~Holdom, J.~Ren and C.~Zhang,
  JHEP {\bf 03} (2015) 028.

\bibitem{Litim:2014uca}
  D.~F.~Litim and F.~Sannino,
  JHEP {\bf 12}  (2014) 178.

\bibitem{QGBH1}
  K.~S.~Stelle,
  Gen.\ Relativ.\ Gravit.\  {\bf 9}, 353 (1978).

\bibitem{QGBH2}
  B.~Holdom,
  Phys.\ Rev.\ D {\bf 66}, 084010 (2002).

\bibitem{QGBH3}
  H.~L¨¹, A.~Perkins, C.~N.~Pope and K.~S.~Stelle,
  Phys.~Rev.~D {\bf 92}, 124019 (2015).

\bibitem{Eardley:2002re}
  D.~M.~Eardley and S.~B.~Giddings,
  Phys.\ Rev.\ D {\bf 66}, 044011 (2002).

\bibitem{Giddings:2004xy}
  S.~B.~Giddings and V.~S.~Rychkov,
  Phys.\ Rev.\ D {\bf 70}, 104026 (2004).

\bibitem{Joshi:2013xoa}
  For a review, see P.~S.~Joshi,
  arXiv:1311.0449.

\bibitem{Torres:2004hk}
  For a review, see D.~F.~Torres and L.~A.~Anchordoqui,
  Rep.\ Prog.\ Phys.\  {\bf 67}, 1663 (2004).

\end{thebibliography}
\end{document}